\documentclass[aps,11pt,superscriptaddress,nofootinbib,notitlepage,showkeys]{revtex4-2}

\pdfoutput=1
\usepackage[T1]{fontenc}
\usepackage{comment}
\usepackage{graphicx}
\usepackage{bm,epsfig,ifthen}
\usepackage{amsfonts,mathtools,amssymb,amsmath,stmaryrd}
\usepackage{color}
\usepackage[breaklinks]{hyperref}
\usepackage{epstopdf, epsfig}
\usepackage{natbib}
\usepackage{grffile}
\usepackage{amsthm}
\usepackage{bbm}
\usepackage{MnSymbol}
\usepackage{algorithm,algorithmic}

\graphicspath{{Figs/}}

\newtheorem{theorem}{Theorem}

\newcommand{\mF}{\mathcal{F}}
\newcommand{\mG}{\mathcal{G}}
\newcommand{\mH}{\mathcal{H}}

\newcommand{\mU}{{\mathcal U}}
\newcommand{\mW}{\mathcal{W}}

\newcommand{\bX}{{\bf X}}

\newcommand{\field}[1]{\mathbb{#1}}
\newcommand{\T}{\boldsymbol{\theta}}

\newcommand{\be}{\begin{equation}}
\newcommand{\ee}{\end{equation}}
\newcommand{\bsp}{\begin{split}}
\newcommand{\esp}{\end{split}}

\newcommand{\bi}{\begin{itemize}}
\newcommand{\ei}{\end{itemize}}

\graphicspath{{figs/}{newfigs/}}

\begin{document}

\title{Non-unique self-similar blowups in {\color{black}shell models}: \\insights from dynamical systems and  machine-learning}

\author{Ciro Campolina}
\affiliation{Calisto, Inria, Universit\'e C\^ote d'Azur, 2004 route des Lucioles, 06902 Sophia Antipolis, France}
\affiliation{Institut de Physique de Nice, Universit\'e C\^ote d'Azur CNRS - UMR 7010, 17 rue Julien Laupr\^etre, 06200 Nice, France}
\author{Eric Simonnet}
\affiliation{Institut de Physique de Nice, Universit\'e C\^ote d'Azur CNRS - UMR 7010, 17 rue Julien Laupr\^etre, 06200 Nice, France}
\author{Simon Thalabard}
\affiliation{Institut de Physique de Nice, Universit\'e C\^ote d'Azur CNRS - UMR 7010, 17 rue Julien Laupr\^etre, 06200 Nice, France}
\email{simon.thalabard@univ-cotedazur.fr}

\date{\today}
\begin{abstract} 
Strong numerical hints exist in favor of a universal  blowup  scenario in the Sabra shell model,
{\color{black}a popular cascade model} of 3D turbulence,   which features complex velocity variables on a geometric progression of scales $\ell_n \propto \lambda ^{-n}$.
 The blowup is thought to be of self-similar type and  characterized by the finite-time convergence towards a  universal profile with  non-Kolmogorov (anomalous) small-scale scaling $\propto \ell_n^{x}$. 
 Solving the underlying nonlinear eigenvalue problem has however proven challenging, and prior insights mainly used the Dombre-Gilson renormalization scheme, transforming self-similar solutions into solitons propagating over infinite rescaled time horizon.
Here, we  further characterize Sabra blowups  by implementing two strategies  targeting  the eigenvalue problem.  The first involves  formal expansion in terms of the bookkeeping parameter  $\delta = (1-x)\log \lambda$, and interpretes the self-similar solution as a (degenerate) homoclinic bifurcation. Using standard bifurcation toolkits, we show that the homoclinic bifurcations identified under finite-truncation of the series converge to the observed Sabra solution. The second strategy uses machine-learning optimization to solve directly for the Sabra eigenvalue. It reveals an intricate phase space, with the presence of a continuous family of non-universal blowup profiles, characterized by various number $N$ of pulses and exponents $x_N\ge x$.
\end{abstract}

\maketitle

\section{introduction}
The blowup problem in  mathematical fluid dynamics addresses the  settlement of non-Lipschitz singularities within a finite-time horizon, out of sufficiently smooth initial data. 
The presence of  blowups and their nature are  believed to be key for
 the  classification of a variety of commonly observed statistical features of turbulent systems, from power laws and multiscaling~\cite{gilson1998two,mailybaev2012computation} to stochastic predictability~\cite{mailybaev2016spontaneous,drivas2021life}. 
With the 3D Euler as a paradigmatic case,
the direct mathematical evidence of blowup from generic initial conditions is however a notoriously hard problem \cite{chae2008incompressible,eggers2008role}, with even direct numerical observations 
leading to conflicting insights \cite{frisch1997there,gibbon2008three,hou2008blowup,bustamante2012interplay,campolina2018chaotic,krstulovic2024initial}.

Presumably, one of the simplest  blowup scenarios is that of asymptotic self-similarity, which describes the finite-time algebraic convergence towards a universal self-similar profile, 
with  specific scaling exponent $x$ prescribed as a non-linear eigenvalue problem \cite{barenblatt1972self,eggers2008role}.  In the 3D Euler setting, an explicit  self-similar axially symmetric solution was recently constructed through machine-learning optimization techniques, pointing towards the relevance of the self-similar scenario  
 at least for certain specific flow configurations \cite{wang2023asymptotic}. 
Self-similar blowups were also identified in 
the simplified setting of  cascade models, including in particular diffusion-approximation (Leith) \cite{leith1967diffusion,nazarenko2011wave} and shell model representations of fluid dynamics \cite{biferale2003shell,nakano1988determination,dombre1998intermittency,andersen2000pulses,katz2005finite,barbato2011energy,mailybaev2012renormalization}.
In  Leith models, self-similar attractors were explicitly identified  in terms of  homoclinic  bifurcations of  low-order but likely intricate dynamical systems \cite{thalabard2015anomalous,galtier2019nonlinear,thalabard2021inverse}, generically prescribing  non-Kolmogorov (anomalous) scaling exponent. In shell models,  following the seminal work by Dombre and Gilson \cite{gilson1998two},  the identification of self-similar blowups relied on numerical simulations exploiting formulations in terms of  similarity variables  \cite{eggers2008role,giga1987characterizing}, which
in essence map self-similar blowups onto travelling waves propagating within an infinite time horizon, whose speed is prescribed by the eigenvalue $x$.
{\color{black}A later work~\cite{l2002quasisolitons} approached the eigenvalue problem by reformulating it as a variation procedure for approximated self-similar solitary peaks (solitons).
Studying local minima of the associated cost functionals, the contribution of trajectories near the self-similar soliton (called quasisolitons) was considered for the asymptotic multiscaling nature of turbulent solutions.}

The purpose of the present paper is to revisit the existence of self-similar blowup solutions in shell models~\cite{biferale2003shell}, by implementing two strategies directly addressing the underlying non-linear eigenvalue problem.
For the sake of clarity, we here focus on the case of the  Sabra shell model~\cite{L1998improved}, known to exhibit self-similar  blowups over a wide range of parameters \cite{mailybaev2012renormalization}.
{\color{black}Still, our results also immediately hold for the GOY model~\cite{gledzer1973system,yamada1987lyapunov}, and our techniques can be extended and applied to other shell models.}
The two distinct strategies provide complementary views on the complexity of the Sabra blowup.
 The first strategy is perturbative and involves a formal expansion in terms of the bookkeeping parameter  $\delta = (1-x)\log \lambda$. Such expansion yields a hierarchy of associated dynamical systems, for which the appearence of a self-similar solution is interpreted as a (degenerate) homoclinic bifurcation. Using standard numerical tools of bifurcation theory, we show that  under any finite-truncation of the expansion series, the homoclinic bifurcation can be tracked down from a classical Hopf bifurcation scenario, with  fast convergence to the observed Sabra solution.
 The second strategy is non-perturbative: it uses machine-learning optimization to solve directly the Sabra eigenvalue problem. This approach reveals an intricate phase space, with the presence of a continuous family of non-universal blowup profiles.

The paper is organized as follows. 
\S \ref{sec:1}  recalls the general Sabra dynamics and its Dombre-Gilson formulation, and formulates the eigenvalue problem.
\S \ref{sec:2} describes the perturbative approach and interpretation of the similarity profile as a global bifurcation.
\S \ref{sec:3} describes the non-perturbative solution using machine learning optimization.
\S \ref{sec:4} formulates concluding remarks.
Some additional material is left to appendices.

\section{Sabra dynamics and the  eigenvalue problem}
\label{sec:1}

\paragraph{Sabra shell model.}
Shell models of turbulence~\cite{biferale2003shell,ditlevsen2010turbulence} are infinite dimensional dynamical systems of continuous time $t \geq 0$, whose variables are complex velocities $u_n(t)$, $n \in \mathbb{Z}$, associated to a geometric progression of scales $\ell_n = k_n^{-1} = \lambda^{-n}$, with $\lambda > 1$ the intershell ratio.
The specific form of the equations is chosen to mimic the spectral Navier-Stokes equations and to keep some of their main properties, like the symmetries and the inviscid invariants of motion.
Many different formulations have been proposed~\cite{gledzer1973system,desnyansky1974evolution,yamada1987lyapunov,ohkitani1989temporal}.
{\color{black}We refer the reader to reference~\cite{plunian2013shell} for a comprehensive list and deduction of shell models in the context of magnetohydrodynamics.}
In this paper, we shall consider the inviscid Sabra model~\cite{L1998improved}, which prescribes a cascade dynamics with nearest and next-nearest shell interactions as
\be
\label{eq:sabra}
\dot u_n = ik_n \left[ \lambda u_{n+2}u_{n+1}^* -(1+c) u_{n+1}u_{n-1}^* - c \lambda^{-1} u_{n-1}u_{n-2} \right] ,\quad n\in \mathbb Z,
\ee
where $i$ is the imaginary unit, and the upper-script stars stand for complex conjugation.
The coefficient $c$ is here prescribed to be negative, and we later write it as $c=-\lambda^{-g}$ with $g$ a positive constant.
In this form, Sabra dynamics exhibits the  two inviscid conserved quantities $E  = \sum_n |u_n|^2$ and $C = \sum_n c^{-n} |u_n|^2=\sum_n (-1)^nk_n^{g} |u_n|^2$.
The first invariant is the kinetic energy, and the second is often called kinetic helicity for not been a sign-defined quantity.
Classical Sabra model is associated to the choice $g = 1$, corresponding to $c=-\lambda^{-1}$.
In this case, the second invariant can be rewritten as $C = \sum_n (-1)^n|u_n\omega_n|$, by introducing the vorticities $\omega_n = ik_nu_n$, and thus $C$ has the same dimension as the full Euler helicity.

Forced and dissipated Sabra shares a number of features with the 3D Navier Stokes equations.
The constant energy flux solution yields Kolmogorov scaling $u_n \propto l_n^{1/3}$, 
and exact inertial laws, akin to Kolmogorov's four-fifths law, also hold~\cite{pisarenko1993further,biferale1998helicity}.
Most importantly,  the structure functions feature  algebraic decay with anomalous exponents ~\cite{jensen1991intermittency,pisarenko1993further,angheluta2006anomalous},   signaling a multifractal behavior similar to the heuristics of  
Navier-Stokes turbulence \cite{L1998improved}. The strong reduction in degrees of freedom however allows for accurate numerical estimations of  arbitrarily high-order structure functions~\cite{de2024extreme, creswell2024anomalous}.

\paragraph{Finite-time blowup.}
The main feature of Sabra that we study in this paper is the developement of singularities in finite time from regular initial data.
By regular data, we mean that the values of $u_n$ decay sufficiently fast with respect to $n$, say the norm $\Vert u \Vert = \left(\sum_{n} k_n^2 |u_n|^2 \right)^{1/2} < \infty$.
In this case, local-in-time well-posedness theorems~\cite{constantin2007regularity} assure that a unique solution exists, and it remains regular over a maximal interval of time $[0,t_\ast)$.
In principle, either the solution keeps the regularity for arbitrarily large times, \textit{i.e.} $t_\ast = \infty$, or there is a singularity at $t_\ast < \infty$, after which the solution cannot be continued in the same regularity class.
In the latter case, we necessarily have the blowup of the norm $\Vert u(t) \Vert \to \infty$ as $t \to t_\ast$.
More specifically, a Beale-Kato-Majda blowup criterion~\cite{constantin2007regularity} shows that $\int_0^{t_\ast} \sup_n |\omega_n|\, dt = \infty$ for the vorticities $\omega_n = ik_nu_n$ in the case of finite-time singularity.
For instance, this implies that if $\sup_n |\omega_n| \sim (t_\ast -t)^{-\beta}$, we must have $\beta \geq 1$.
While the finite-time blowup has been proved for other shell model formulations~\cite{katz2005finite,cheskidov2008blow}, there are no rigorous statements about whether $t_\ast$ is finite or not for Sabra.
Nevertheless, all previous numerical works~\cite{dombre1998intermittency,mailybaev2012renormalization,mailybaev2013blowup,mailybaev2016spontaneous,mailybaev2016spontaneously} attest the existence of a finite-time singularity $t_\ast < \infty$.

\begin{figure}
	\includegraphics[width=\textwidth]{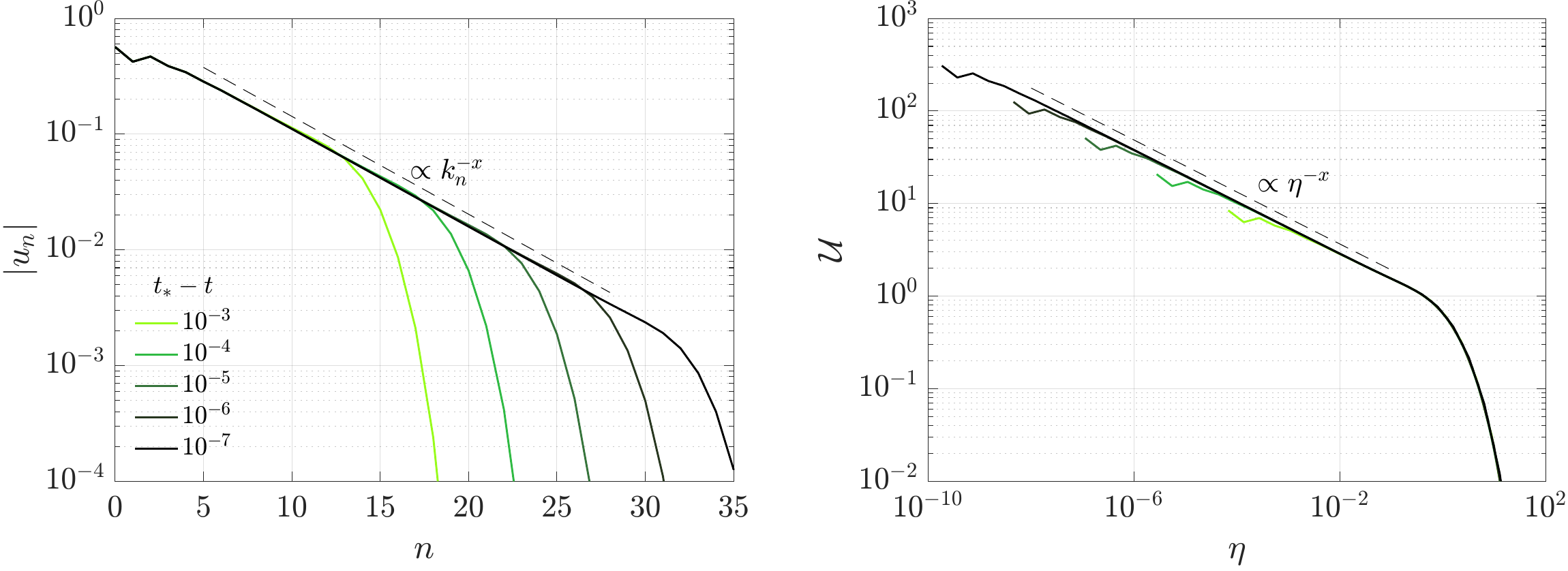}
	\caption{Direct numerical integrations of classical Sabra dynamics~\eqref{eq:sabra}, $\lambda = 2$ and $c = -\lambda^{-1}$.
		For the computations, we simulate only the points $n \geq 0$, and we use the boundary conditions $u_{-1} = u_{-2} = 0$.
		The initial data is nonzero only for the two first points $u_0 = -i$ and $u_1 = -0.5i$.
		Left: Shell velocities $|u_n|$ in log scale as function of $n$ for different instants of time approaching the blowup time $t_\ast \approx 1.243$.
		The solution develops a power law $|u_n| \propto k_n^{-x}$ with $x \approx 0.281$.
		Right: The velocity self-similar profile $\mU(\eta)$ computed from the numerical integrations using formula~\eqref{eq:self}.
		Colors correspond to different instants of time.
		The profile displays the infra-red power law $\mU \propto \eta^{-x}$ and the ultra-violet sharp front.}
	\label{fig:1}
\end{figure}

Fig.\ref{fig:1} illustrates an archetypical numerical integration of the classical Sabra model.
The solution initially restricted to small $n$ extends to ultraviolet scales by developing the power law $|u_n| \propto k_n^{-x}$ in the finite time $t_\ast \approx 1.243$.
The scaling exponent is empirically estimated as $x \approx 0.281$, and it was previously found in literature~\cite{mailybaev2012renormalization}.
Numerical computations also reveal the asymptotics $\Vert u \Vert \sim (t_\ast-t)^{-1}$ and $\sup_n |\omega_n| \sim (t_\ast - t)^{-1}$ as $t \to t_\ast$, which are compatible with blowup criteria.
Most importantly, one verifies that the shape of the spectra front extending from infrared to ultraviolet is preserved up to rescaling, suggesting self-similar dynamics.

\begin{figure}
	\includegraphics[width=\textwidth]{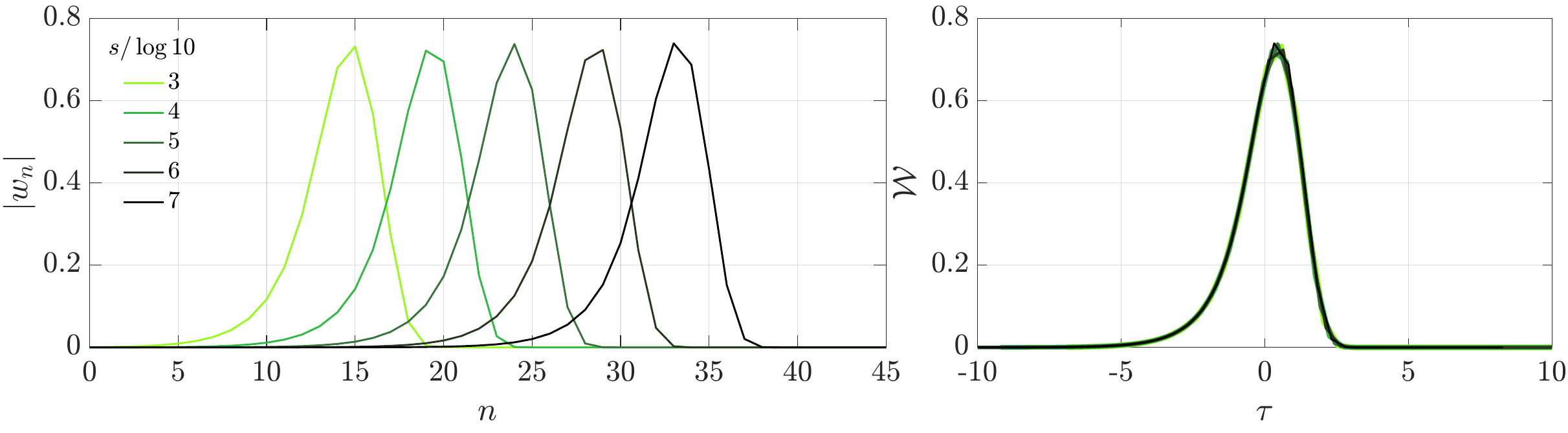}
	\caption{Left:~Sabra dynamics for the Dombre-Gilson renormalized vorticities $|w_n|$ as function of $n$ for different logarithmic times $s$, which correspond exactly to the instants of time $t$ displayed in Fig.~\ref{fig:1}.
	Right:~The homogeneous vorticity self-similar profile $\mW(\tau)$ computed from the direct numerical simulations using formula~\eqref{eq:wtau}.}
	\label{fig:2}
\end{figure}

\paragraph{Dombre-Gilson renormalization.}
Self-similarity is better characterized when considering the renormalization scheme introduced by Dombre and Gilson (DG)~\cite{dombre1998intermittency}.
Strictly speaking, the DG scheme introduces the logarithmic time and the rescaled vorticities
\be
\label{eq:dg}
t_*-t  \mapsto  s = - \log (t_*-t),\quad u_n \mapsto w_n=i  k_n u_n e^{- s}.
\ee
Such transformation maps the finite-time blowup to an infinite-horizon essentially-bounded dynamics.
Sabra equations~\eqref{eq:sabra} are then written in those new variables as
\be
\label{eq:sabra2}
\dfrac{dw_n}{ds} = 
- w_n   -  \left[ \lambda^{-2} w_{n+2}w_{n+1}^* -(1+c) w_{n+1}w_{n-1}^* +c \lambda^2  w_{n-1}w_{n-2} \right].
\ee

\begin{figure}
	\includegraphics[width=0.8\textwidth]{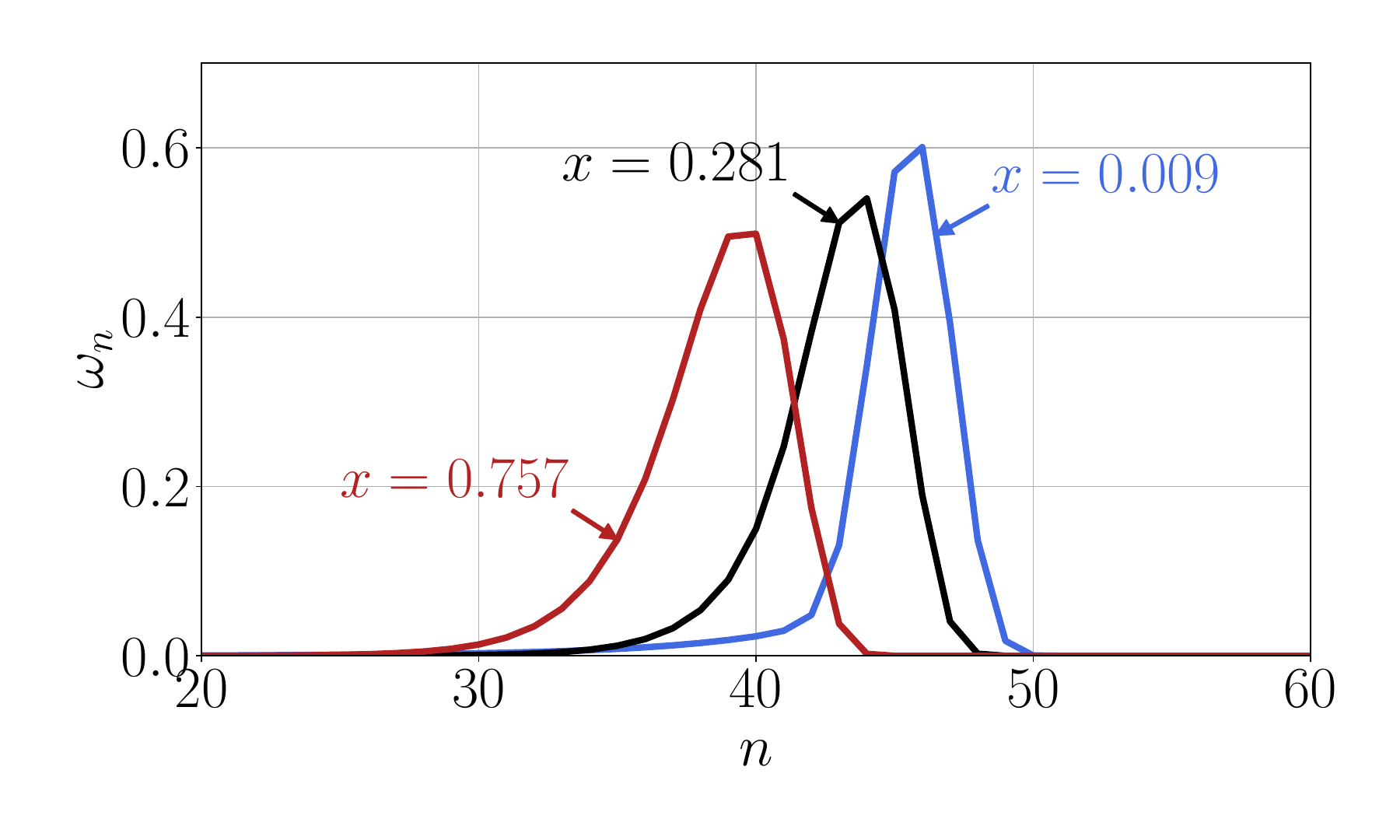}
	\caption{Typical blow-up profiles obtained by integrating~\eqref{eq:sabra2} in log scale and renormalized to
		have unit $\ell^2$ norm.
		The DG solutions are shown as a function of $n$ for some fixed time well after transients.
		All these solutions are indeed travelling waves.
		Three cases are shown:  the red curve corresponds to $(\lambda,g) = (7.389,0.2)$, the black curve is for $(\lambda,g) = (2,1)$,
		and the blue curve is $(\lambda,g) = (1.221,2)$.
		}
	\label{fig:3}
\end{figure}

Numerical simulations show that the self-similar blowups in the original Sabra equations correspond to traveling wave solutions in these new renormalized variables~\cite{dombre1998intermittency,mailybaev2012renormalization}.
Fig.~\ref{fig:2} illustrates such traveling waves.
The scaling exponent $x$ of the blowup solution depends upon the parameters of the problem, namely the intershell ratio $\lambda$ and the $g$ parameter prescribing the second inviscid invariant.
More interestingly, the wave profile may vary for different parameters.
Numerical experiments suggest the existence of at least one traveling wave for each pair of values $(\lambda,g)$.
Fig.~\ref{fig:3} compares distinct solutions obtained from three diferent choices of parameters.

Self-similarity is related to the fact that the waves preserve exactly the profile in their constant speed propagation, while the speed of propagation itself can be related to the scaling exponent $x$ (see~\cite{mailybaev2016spontaneously} for more details).
The wave profile is recovered by considering a Poincaré map defined over a set of successive discrete instants of renormalized time $s_n$ for which the center of mass of the wave lies on the nodes $n$.
In this case, the self-similar blowup corresponds to a fixed-point attractor of the Poincaré map, which is precisely the wave profile.
Using this approach to other shell models reveals rather intricate blowup scenarios, like periodic, quasi-periodic and chaotic blowups~\cite{mailybaev2013bifurcations}.
Nevertheless, the blowup of Sabra was verified to be self-similar for a wide range of parameters.
It is even conjectured that the blowup in Sabra might be universal, as a manifestation of the renormalization group formalism~\cite{mailybaev2012renormalization}.
In dynamical systems terminology, this would correspond to a fixed point attractor with non-zero measure basin of attraction.
These particular statements are in the core of our investigations on the next sections.
Particularly, we shall support the existence of multiple unstable self-similar blowups in Sabra, which have not been yet reported in literature, up to our knowledge.
This reveals new insights about the important problem of universality in Sabra finite-time blowup.

\paragraph{Sabra self-similarity eigenvalue problem.}
While the DG scheme proves a very useful tool to classify the  different types of blowups and to track their bifurcations in general shell models, here we focus only on the self-similar scenario in Sabra.
In this case, the usual and most effective approach is to deduce governing equations directly for the self-similar profile~\cite{barenblatt1996scaling}, which we do now.
The strategy is to write a general ansatz for the profile together with the proper boundary conditions.
Then we undergo some variable transformations to make the system autonomous.
Such procedure  leads to a nonlinear eigenvalue problem, whose eigenvector is the renormalized self-similar profile, and the eigenvalue is the scaling exponent $x$.
For simplicity, we here focus on the case of purely imaginary variables $u_n \in i\mathbb{R}$.
{\color{black}For this reason, our results equaly apply to the GOY shell model~\cite{gledzer1973system,yamada1987lyapunov}, which differs from Sabra only on the complex conjugations of the nonlinear term.}
Specifically, we consider nontrivial solutions of the form
\be
\label{eq:self}
u_n(t) = -i (t_\ast-t)^{\frac{x}{1-x}} \mU(\eta_n), \quad \text{for} \ \eta_n = (t_\ast-t)^{\frac{1}{1-x}}k_n.
\ee
The minus sign is conventional.
The real-valued function $\mU(\eta)$ is the self-similar profile, which couples the infrared power-law scaling
\be
\label{eq:BCinfrared}
\mU \propto \eta^{-x} \quad \text{as} \ \eta \to 0,
\ee
to a sharp ultraviolet front
\be
\label{eq:BCultraviolet}
\mU \to 0 \ \text{faster than any power law as} \ \eta \to \infty.
\ee
We draw the self-similar profile $\mU(\eta)$ from direct numerical integrations in Fig.~\ref{fig:1}.
Ansatz~\eqref{eq:self} and boundary conditions~\eqref{eq:BCinfrared}-\eqref{eq:BCultraviolet} are motivated by the scaling symmetry of Sabra and the empirical evidence of an infrared power-law scaling $|u_n| \propto k_n^{-x}$ extending up to an ultraviolet sharp front---see Fig.~\ref{fig:1} for the numerical picture.
We remark that the  scaling exponent  is necessarily bounded above, $x < 1$. With this bound,  Eq.~\eqref{eq:self} describes  a front $k_*\propto (t_*-t)^{-1/(1-x)}$ propagating to the UV in finite-time  as $t \to t_*$.
Under the ansatz \eqref{eq:self}, Sabra dynamics~\eqref{eq:sabra} prescribes the equation for the profile $\mU(\eta)$ as
\be
\begin{split}
	x\, \mU (\eta_n)+ \eta_n\, \mU'(\eta_n)   = 
	(1-x)\eta_n
	\left[
	\lambda\, \mU\left(\eta_{n+2}\right)\,\mU\left(\eta_{n+1}\right)
	\hspace{-0.3cm}
	\phantom{\dfrac{1}{\lambda}}  \right.
	& \left.-\left(1+c\right)\, \mU\left(\eta_{n+1}\right)\,
	\mU\left(\eta_{n-1}\right) 
	+
	\right.\\
	& 
	\left.
	\phantom{\dfrac{1}{\lambda}}
	+ c \lambda^{-1} \mU\left(\eta_{n-1}\right)\,\mU\left(\eta_{n-2}\right)\right].
\end{split}
\label{eq:BVa}
\ee
Together with the  boundary conditions \eqref{eq:BCinfrared}-\eqref{eq:BCultraviolet}, this is the fundamental boundary value problem (BVP) for the self-similar blowup profile $\mU(\eta)$.

For further analysis, it proves convenient to recast Eq.~\eqref{eq:BVa} into an autonomous form.
We achieve this upon introducing the logarithmic similarity variable and the rescaled vorticity profile
\be
\label{eq:wtau}
\tau_n = \log \eta_n,\quad \mW(\tau_n) = e^{\tau_n}\, \mU\left(e^{\tau_n}\right).
\ee
After rescaling the new variable $\tau_n \mapsto \tau_n/(1-x)$, the Eq.~\eqref{eq:BVa} becomes the autonomous dynamical system
\be
\label{eq:wtaun}
\mW'(\tau_n) = \mW(\tau_n) + \lambda^{-2}\mW(\tau_{n+2})\mW(\tau_{n+1}) - (1+c)\mW(\tau_{n+1})\mW(\tau_{n-1}) + c\lambda^2 \mW(\tau_{n-1})\mW(\tau_{n-2}),
\ee
subject to the boundary conditions
\be
\label{eq:BVc}
\mW \underset{-\infty}{\sim} e^\tau \to 0 \quad  \text{and} \quad \mW \underset\infty{\to} 0.
\ee
Observe that the logarithmic transformation~\eqref{eq:wtau} maps shifts in $n$ into additions to $\tau_n$ as
\be
\tau_{n+j} = \tau_n + j \delta, \quad j \in \mathbb{Z},
\ee
for the parameter
\be
\label{eq:delta}
\delta = (1-x)\log \lambda.
\ee
Fig.~\ref{fig:2} plots the profile $\mW$ computed from direct numerical integrations of Sabra.

In principle, Sabra dynamics prescribes the profile $\mW$ only at the discrete scales $\tau_n$.
This poses difficulties, as the left-hand side of~\eqref{eq:wtaun} involves evaluating the derivative $\mW'$.
To analyze the Sabra BVP, we therefore extend the problem to the full line, \textit{i.e.}, we consider the profile $\mW(\tau)$ for the continuous variable $\tau \in \mathbb{R}$.
We remark that this is well justified since such continuous limit does not require estimating differential operators.
Finally, the Sabra BVP becomes the functional fixed-point problem
\be
\label{eq:BVb}
\dot \mW  = \mathcal F\left[ \mW \right],\quad   \mathcal F  \left[ \mW \right] =
\mW+ \lambda^{-2}\, \mW_{2\delta}\,\mW_{\delta} -(1+c)\, \mW_{\delta}\,\mW_{-\delta} + c \lambda^2 \mW_{-\delta}\,\mW_{-2\delta},
\ee
with the shorthand $\mW_{j\delta} (\tau) = \mW(\tau + j\delta)$, $j \in \mathbb{Z}$, the parameter $\delta$ given by~\eqref{eq:delta}, and the boundary conditions~\eqref{eq:BVc}.
Eq.~\eqref{eq:BVb} can also be interpreted as a nonlinear eigenvalue problem, whose eigenvector is the renormalized self-similar profile $\mW$, and the associated eigenvalue is the scaling exponent~$x$.
This is the main problem we shall analyze in this paper.

\begin{figure}
\hspace*{-0.2cm}\includegraphics[width=0.6\textwidth]{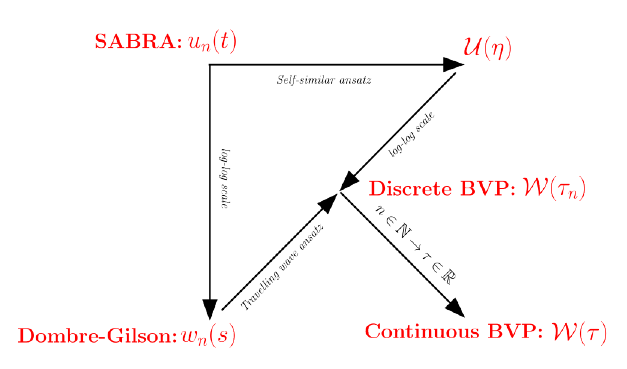}
	\caption{Starting from Sabra shell model~\eqref{eq:sabra}: correspondence between Dombre-Gilson~\eqref{eq:sabra2} and our BVP problem~\eqref{eq:BVb}.}
	\label{fig:4}
\end{figure}

Some remarks should be addressed.
We point out that the solution of the autonomous BVP gives the profile of the traveling wave obtained from the DG scheme.
Indeed, the dynamics featured in Eq.~\eqref{eq:BVb} is recovered from DG variables~\eqref{eq:sabra2} by considering the travelling wave solution
\be
w_n(\tau)  = \mW\left( \tau_n \right), \quad \tau_n=(1-x) \log \lambda  \left(n-n_*\right), \quad n_*(\tau)= \dfrac{\tau }{(1-x) \log \lambda},
\ee
and identifying $w_{n+j}(\tau) = \mW\left(\tau_{n+j}\right) = \mW_j (\tau_n)$---recalling the shorthand  $\mW_{j\delta}(\tau) = \mW\left( \tau + j \delta \right)$.
As such, the DG rescaling scheme, be it obtained from the change of variables  \eqref{eq:wtau} or 	from \eqref{eq:dg}, maps the similarity ansatz onto a traveling wave solution $\mW(\tau_n)$, which propagates in shell space over the infinite time $\tau$ horizon.
In the end, the final BVP problem~\eqref{eq:BVb} emerges as a continuous limit {\color{black}(i.e. taking $\tau_n, \ n \in \mathbb{N} \mapsto \tau \in \mathbb{R}$)} of either a self-similar ansatz or a Dombre-Gilson traveling wave solution.
Fig.~\ref{fig:4} illustrates the correspondence between the different approaches.

From a topological viewpoint, the origin~$\mW \equiv 0$ is a fixed point of $\mathcal F$.
The vanishing boundary conditions~\eqref{eq:BVc} at $\tau \to \pm \infty$ imply that the similarity profile $\mW$ is obtained as a homoclinic trajectory of the homogeneous functional dynamics~\eqref{eq:BVb}.
At this point, the difficulty lies in the fact that~\eqref{eq:BVb} is not a classical delay differential equation (see \textit{e.g.}~\cite{diekmann2012delay}), but rather a general Riccati-like functional equation mixing positive and negative time shifts.
We are not aware of any general theory for the analysis of such kind of system.

Our main goal in the next sections is to study existence, uniqueness and stability of blowup solutions in Sabra through the autonomous BVP~\eqref{eq:BVb}.
In the absence of general analytical theories, we approach this problem using  two distinct strategies.
The first strategy consists in Taylor-expanding the shifted functions $\mW_{j\delta}$ with respect to the bookkeeping parameter $\delta$, and then use dynamical systems tools.
This perturbative approach is reported in~\S\ref{sec:2}.
The second strategy consists in solving directly the problem without any \textit{a priori} simplification.
For this, we use machine learning tools.
We address this non-perturbative approach in~\S\ref{sec:3}.

\section{Perturbative analysis \& Sabra hierarchy }
\label{sec:2}

The main difficulty of the similarity equation resides on the field time shifts.
To overcome this, we consider a systematic perturbative approach.
We undertake Taylor expansions of the shifted profiles $\mW_{j\delta}$ in Eq.~\eqref{eq:BVb} with respect to the bookkeeping parameter $\delta$.
This yields a hierarchy of dynamical systems whose bifurcations describe Sabra blowup solutions.
We start by the local analysis of the main fixed point equation.

\paragraph{Fixed points and Hopf bifurcation.}
Local analysis of Sabra BVP~\eqref{eq:BVb} identifies two elementary fixed points $\mF\left[\mW\right]=0$ , namely the  constant profiles 
\be
\label{eq:constant}
\mW_0= 0 \quad \text{and} \quad \mW_{\rm H} =\frac{\lambda^2}{(\lambda^2-1) (1-c \lambda^2)} > 0.
\ee
Direct calculations reveal that $\mW_0$ behaves like a saddle.
We are interested in studying the stability of $\mW_{\rm H}$ when $x$ varies.
For this, we are able to show rigorously the following result, that we state as a theorem.
\begin{theorem}\label{thm:1}
Assume $c < 0$, and $\lambda > 1$.
The fixed point $\mW_{\rm H}$ loses its stability through an Hopf bifurcation for the critical exponent $x_{\rm Hopf}$
\be \label{hopfformula}
x_{\rm Hopf}= 1 -    \frac{(\lambda^2-1)(1-c \lambda^2)}{(1-c \lambda^4) \log \lambda} \frac{\arccos \rho}{(1+2\rho)
\sqrt{1-\rho^2}} < 1,
\ee
with frequency
\be
\xi_{\rm Hopf} = \frac{1-c \lambda^4}{(\lambda^2-1)(1-c \lambda^2)} (1+2\rho) \sqrt{1-\rho^2},
\ee
where $\rho$ is the unique root $|\rho| \leq 1$ of $\rho^2 + (\frac12-R) \rho + \frac12 R -1 = 0$, with $R = \frac{c+1}{c \lambda^2 + \lambda^{-2}}$. 
The solution is stable when $x > x_{\rm Hopf}$ and unstable otherwise.
\end{theorem}

The proof is given in the Appendix~\ref{app:local}.
The remarkable aspect of this result is that it applies for all values of $c < 0$ and $\lambda > 1$.  Another striking fact is that, in the limit
$\lambda \to 1^+$, the Hopf bifurcation parameters $x_{\rm Hopf}$ have a well-defined finite limit, with the frequecies $\xi_{\rm Hopf}$ growing to infinity.

System~\eqref{eq:BVb} appears to always follow the same bifurcation sequence when $x$ decreases:
\begin{center}
{\it stable fixed point $\to$ Hopf bifurcation $\to$ stable limit cycle $\to$ homoclinic orbit.}
\end{center} 
However, in its continuous functional form, the main difficulty in the analysis of the Sabra BVP~\eqref{eq:BVb} comes from the presence of the $x$-dependent time shifts through integer multiples of $\delta = (1-x)\log\lambda$.
This constitutes a class of dynamical systems for which (infinite-dimensional) bifurcation theory might be non-standard.

\paragraph{The Sabra hierarchy. }
 The idea of the  perturbative approach is to consider expansions of the Sabra BVP in terms  on the bookkeeping parameter $\delta=  (1-x) \log \lambda$, in order to interprete its solution as a global (homoclinic) bifurcation arising in  an infinite-dimensional dynamical system.
 We show below that this system can be truncated and analyzed  with standard dynamical system tools at any finite order $M$. 

Explicitly, when truncated at order $O(\delta^M)$, the Sabra  BVP \eqref{eq:BVb} maps to the dynamical system
\be
\label{eq:BVPm}
\left\lbrace
\begin{split}
	 X_k'(\theta) & = X_0X_{k+1}, \quad ( 0\le k \le M-2),\quad \quad  X'_{M-1}(\theta)  = \dfrac{1}{\sigma_{0M}}\;\mG_M\left[\bX\right],\\
 	\text{with } &   \mG_M\left[\bX\right]= - X_0 + \dfrac{X_1}{\delta}  -\dfrac{1}{2}\sum_{k=0}^{M-1} \sum_{\substack{i + j= k \\ 0 \le i,j \le k}}  \sigma_{ij} X_i X_{j}-\dfrac{1}{2} \sum_{\substack{i+j=  M\\ 1 \le i,j \le M-1}}  \sigma_{ij} X_i X_{j},
\end{split}
\right.
\ee
It is formulated in terms of the renormalized time $\theta$ and the rescaled vorticity derivatives $\bX$ as
\be
	\label{eq:theta}
	\bX = (X_0 \dots X_{M-1}),\quad X_k(\theta) = \delta^k \mW^{(k)}(\tau),\quad \theta = \dfrac{1}{\delta}\int_0^\tau \dfrac{d\tau'}{\mW(\tau')},
\ee
and featuring the symmetric coefficients
\be
\label{eq:coeffs}
\sigma_{ij} = \frac{\lambda^{-2} (2^i+2^j)}{i!j!} \left( (-1)^{i+j} c \lambda^4 - \frac{(-1)^i+(-1)^j}{2^i+2^j} (1+c)\lambda^2
+ 1
\right).
\ee
Besides, the boundary conditions~\eqref{eq:BVc} prescribe
\be
\bX \to 0 \quad \text{as} \quad \theta \to \pm \infty
\ee
for the similarity solutions.
We refer the reader to the Appendix~\ref{app:hierarchy} for a detailed derivation of~\eqref{eq:BVPm}--\eqref{eq:coeffs}.
System~\eqref{eq:BVPm} defines  a hierarchy  of dynamical systems of increasing dimension $M$, which formally retrieves the Sabra BVP~\eqref{eq:BVb} when $M\to \infty$, and which we later refer to as the Sabra hierarchy.
The low-order systems can be explicitly given as
\be
\label{eq:Sabra234}
\left\lbrace
\begin{split}
	M=2 \quad  X_0'& = X_0X_{1},\; \sigma_{02} X_1' =- X_0 + \dfrac{X_1}{\delta}  - \dfrac{1}{2}\sigma_{00} X_0^2- \sigma_{01} X_0X_1-\dfrac{1}{2}   \sigma_{11} X_1^2,\\
	M=3 \quad  X_0'& = X_0X_{1},\;   X_1' = X_0X_{2}, \;\sigma_{03} X_2' =G_2-\sigma_{02}X_0X_2 -\sigma_{12}X_1X_2 \\
	M=4 \quad  X_0'& = X_0X_{1},\;   X_1' = X_0X_{2},\; X_2' = X_0X_{3}, \; \sigma_{04} X_3' =G_3-\sigma_{03}X_0X_3- \sigma_{13}X_1X_3 -\dfrac{\sigma_{22}}{2}X_2^2 \\
\end{split}
\right.
\ee
and so on, where we have used the shorthand $G_M = \mathcal G_M[\bX]=\sigma_{0M}X'_{M-1}$, satisfying
\be
 G_{m+1}=G_m  - \sigma_{0M} X_0 X_{M}-\dfrac{1}{2} \sum_{\substack{i+j=  M+1\\ 1 \le i,j \le M}}  \sigma_{ij} X_i X_{j}.
\ee

\paragraph{The homoclinic explosion.}
\begin{figure}
\includegraphics[width=0.49\textwidth]{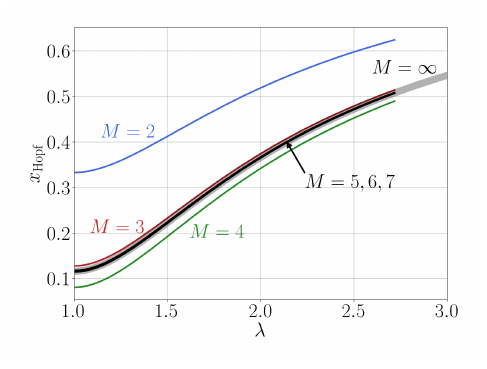}
\includegraphics[width=0.49\textwidth]{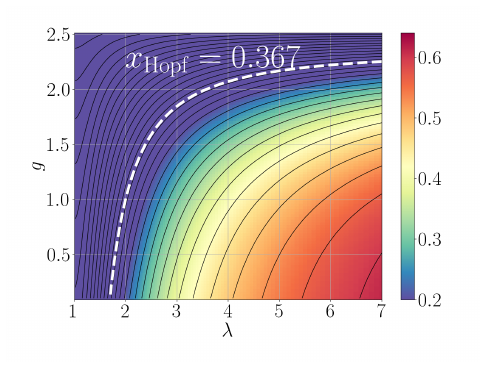}
\caption{Left:~Hopf bifurcation parameters $x_{\rm Hopf}$ as a function of $\lambda$, for $M$ from $2$ to $7$. 
Here, $g=1$ is kept fixed. The value $M = \infty$ is computed from the exact  formula~\eqref{hopfformula}.
		Right:~Hopf parameters in in the plane $( \lambda, g )$ for $M = \infty$ using the exact formula~\eqref{hopfformula}.
		The white curve indicates  the reference case $\lambda=2$, $g=1$. Note that the $\lambda$ range is different in the left and right panels.
}
\label{fig:5}
\end{figure}

\begin{figure}
\begin{minipage}{0.49\textwidth}
\includegraphics[width=\textwidth]{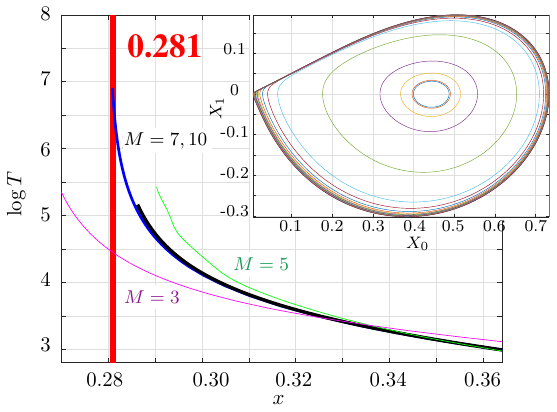}
\end{minipage}
\begin{minipage}{0.49\textwidth}
\includegraphics[width=\textwidth]{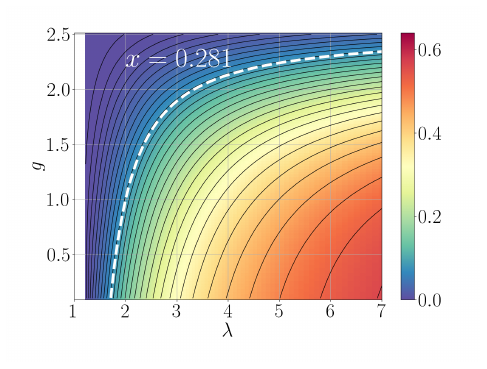}
\end{minipage}
\caption{Left:~Homoclinic explosion showing period $T$ \textit{vs} $x$ for various $M$,
and reference  parameters $\lambda = 2$ and $g = 1$.
	Inset shows the limit cycles tracked for $M=7$.
	Right:~Critical $x$ in Sabra (\ref{eq:sabra}) as a function of both $\lambda$ and $g$ by direct computations.
	The white curve corresponds to the reference case with $g=1$, $\lambda=2$.
}
\label{fig:6}
\end{figure}

The hierarchy~\eqref{eq:BVPm} captures the homoclinic bifurcation relevant for the similarity profiles observed in Sabra direct numerical simulations.
Local analysis reveals that all the truncated systems have the same ($M$-independent) fixed points than~\eqref{eq:BVb}, namely the origin $\bX_0 = (0,0,\dots)$ and $\bX_{\rm H}=(\mW_{\rm H},0,0,\dots)$ --- see Expression~\eqref{eq:constant} of $\mW_{\rm H}$.
However, only the point $\bX_{\rm H}$ is necessarily isolated.
The origin is part of a wider set  
\be
\label{eq:Z0}
\mathcal Z_0^{(M)} = \left\lbrace \bX=(0,\bX_+), \text{ with } \bX_+ \in \mathbb R^{M-1} \; \& \; \mG_M\left[(0,\bX_+)\right] = 0\right\rbrace.
\ee
This structure gives rise to a classical co-dimension 1 scenario of homoclinic bifurcation, which we identify computationally using standard bifurcation tracking software (PyCont and BifurcationToolkit \cite{veltz:hal-02902346, clewley2007pyds}).

For each $M$, the isolated fixed point $\bX_{\rm H}$ undergoes a Hopf bifurcation  at $x_{\rm Hopf}^{(M)}$.
The resulting limit cycle then collides with the invariant set $\mathcal Z_0^{(M)}$ at $x=x_\star^{(M)}$.
The critical values $x_{\rm Hopf}^{(M)}$ and $x_\star^{(M)}$ depend upon the order $M$.
Nevertheless, in the limit $M \to +\infty$, the value $x_{\rm Hopf}^{(M)}$ converges towards the theoretical parameter $x_{\rm Hopf}$ from Theorem~\ref{thm:1}, and $x_\star^{(M)}$ converges to the numerically observed exponent $x^\star$.
We show this in Figs.~\ref{fig:5} and~\ref{fig:6}.
In Fig.~\ref{fig:5}, we plot the Hopf bifurcations for different Sabra parameters $\lambda$ and $g$.
Particularly, one observes a very fast convergence of the truncated parameters $x_{\rm Hopf}^{(M)}$ to the theoretical one $x_{\rm Hopf}$ with increasing $M$.
The curves are visually indistinguishable for $M = 5,6,7$.
In Fig.~\ref{fig:6}, we illustrate the homoclinic explosion, \textit{i.e.} the continuous evolution from the limit cycle to the homoclinic orbit as we vary $x$ from $x_{\rm Hopf}^{(M)}$ to $x_\star^{(M)}$.
The homoclinic parameter $x_\star^{(M)}$ also quickly converges to the numerically observed values $x^\star$.
The fast convergence suggests that, although of infinite dimensionality, the Sabra blowup is accurately described by a low-order model. 

\begin{figure}
\includegraphics[width=0.49\textwidth]{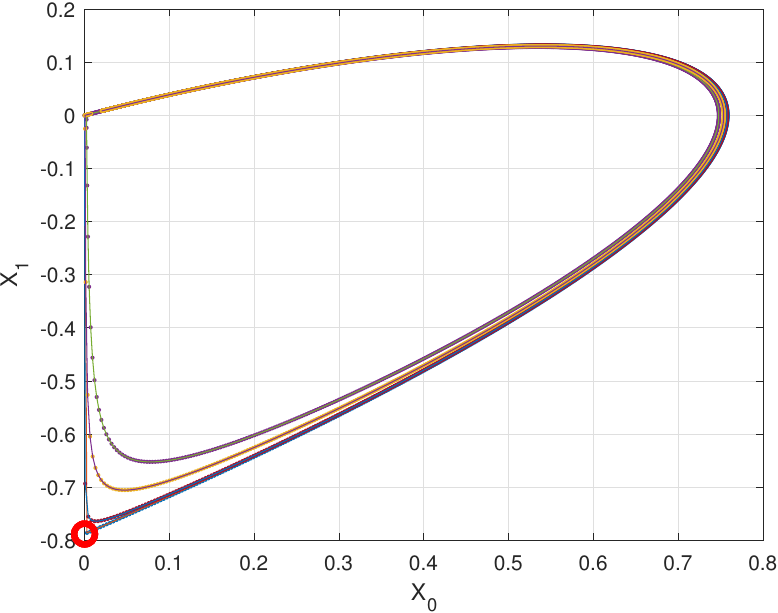}
\includegraphics[width=0.49\textwidth]{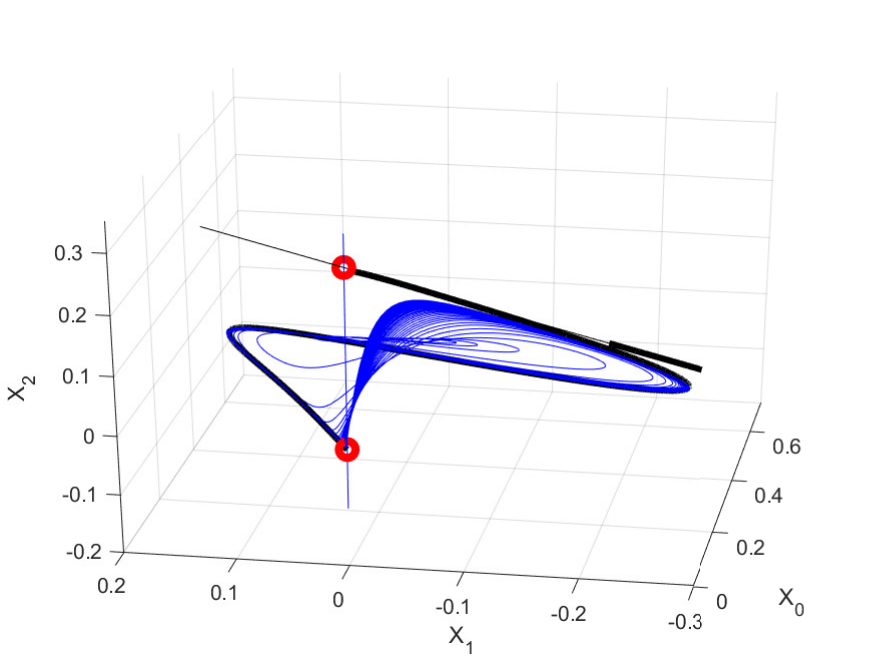}
\caption{Degenerate homoclinic orbits for $M=2$ (left) and $M=3$ (right), resulting from the collision of the limit cycles with the continuum set $ \mathcal Z_0^{(M)}$.}
\label{fig:7}
\end{figure}

\paragraph{Degeneracies.}
At any order, the Sabra hierarchy~\eqref{eq:BVPm} exhibits a continuous set of fixed points: This type of  degeneracy is  inherited  from the singular nature of the Sabra BVP, and  also appears in Leith models \cite{grebenev2013self, thalabard2015anomalous, thalabard2021inverse}. The degeneracy is removed through the nonlinear time change \eqref{eq:theta} but signals that  the homoclinic explosion might be degenerate. 
This is better grasped by considering the low-order models of the Sabra hierarchy, prescribed by Eqs.~\eqref{eq:Sabra234}.
For $M=2$, the set of fixed points $\mathcal Z_0^{(2)}$ is the two-point set
\be
\mathcal Z_0^{(2)} = \{ {\bf 0} \} \cup \{ \bX_c \},\quad \bX^{(2)}_c= \left(0, \dfrac2{\sigma_{11} \delta}\right),\quad \sigma_{11} = 4 \lambda^{-2} \left( c \lambda^4 +\frac12 (1+c) \lambda^2 + 1 \right),
\ee
which for the particular shell ratio $\lambda=2$ yields $\bX_c \simeq (0,-0.78)$.
The left panel of Fig.~\ref{fig:7} shows that as $x$ approaches $x_*^{(2)} \approx 0.389$, the limit cycle does not collide with the origin, but rather with $\bX_c^{(2)}$.
This suggests that the homoclinic cycle is made of two heteroclinic components, namely $0 \to  \bX^{(2)}_{c}$ and $  \bX^{(2)}_{c} \to 0$.
The same type of scenario is seen explicitly  for the case $M=3$.
The set of fixed points $\mathcal Z_0^{(3)}$ identifies with the union of two lines, that is 
\be
	X_0=0,  X_1=0, X_2 \in \mathbb R \quad \text{or} \quad X_0=0,  X_1 \in \mathbb R ,\quad \sigma_{12} X_2  =  -\sigma_{11}X_1 +\dfrac{ 1}{\delta}=0.
\ee
Fig. \ref{fig:7} (right) suggests  that the limit cycles intersect $\mathcal Z_0^{(3)}$ at the point
$\bX_c^{(3)} = \left(0,0,1/(\delta \sigma_{12})\right)$. 

\section{Non-perturbative approach using machine-learning}
\label{sec:3}

In the previous section, we showed that the perturbative approach captures the main numerically observed blowup features.
We want to attest that those observations are indeed properties of the original BVP, and not only of the truncated systems.
Therefore, we would like to directly solve the functional equation~\eqref{eq:BVb} without going through any perturbative approximation.
One cannot rely on a time marching approach since both future and past can influence the state at $\tau$.
A simple idea is therefore to formulate~\eqref{eq:BVb} as an optimization problem.
{\color{black}A similar idea was considered in~\cite{l2002quasisolitons}, where a variation principle was formulated for self-similar solitary peaks.
There, such solitons were approximated by rational trial functions.}
In the present work, we  parametrize the solution by a deep neural network (NN), an approach in the spirit of the so-called Physics-Informed-Neural-Networks (PINNs) method~\cite{Lagaris,pinns}.
In the end, we show that {\color{red}our} approach enlarges our understanding of the problem: It  recovers the self-similar blowups previously discussed, but also uncovers new similarity profiles, which to our knowledge were not reported before.

The problem consists in finding $(x,\mW)$ such that $\dot \mW = \mF[\mW]$, where $x$ appears in $\mF$ only in the delays, but not in the coefficients.
Therefore, it is convenient to understand $\mW$ as a function of both $\tau$ and $x$.
Let us consider the following cost functional
\be\label{mlcost}
{\cal C}[\mW] = \int_{\field{R}} \int_{x_{\rm min}}^{x_{\rm max}} \left| \dot \mW - \mF[\mW] \right|^2~~dx~d\tau.
\ee
The corresponding optimization problem formulates as finding the argmin solution, called $\mW^\star$,
\be
\mW^\star(\tau,x) = \underset{\mW \in V}{\rm argmin}~{\cal C}[\mW] 
\ee
within the functional space
\be
V = \{\mW:(\tau,x) \mapsto \mW(\tau,x) ~~|~~
\mW(\tau,x) \underset{\tau \to -\infty}{\sim} C_{-\infty} e^\tau  ~ \text{and} ~\lim_{\tau \to +\infty} \mW(\tau,x) = 0 \},
\ee
whose functions satisfy the problem boundary conditions.
Here, $C_{-\infty}$ is the real prefactor of the infrared asymptotics, which will be important in the computational formulation.
In this approach, we actually obtain a one-parameter family of solutions indexed by $x$.
Then, we compute for which $x$ the fixed point problem is exactly satisfied.
The main difficulty is to prescribe the boundary conditions, \textit{i.e.} to guarantee that $\mW \in V$.
We explain the details next.

\paragraph{Neural Network parametrization and ansatz.}
First of all, we affirm that we can set the prefactor $C_{-\infty}$ in the definition of $V$ to be a fixed nonzero constant.
Indeed, the equations are invariant to time translations.
Therefore, fixing a value is equivalent to choosing some time reference frame.
Moreover, a nonzero value filters out the trivial identically zero solution.
Hereafter we set $C_{-\infty} = 1$.
Next, one must be carefull in the representation of the whole $\tau$ line.
In practice, we choose $\tau \in [-T,T]$ for $T \gg 1$.
We then consider the following neural ansatz
\be
\label{nnansatz}
\mW_{\rm nn}(\tau,x;\T) = \chi_-(\tau) \chi_+(\tau) (1 + {\cal N}( \tau, x;\T))^2, ~~\chi_{\pm}(\tau) = \left(1+{\rm e}^{-(\tau/2 \pm\sigma)}\right)^{-1},
\ee
where ${\cal N}$ is a simple fully-connected feedforward (or perceptron) neural network, \textit{i.e.} ${\cal N}(\tau,x;\T) = ({\cal N}_{\rm out} \circ {\cal N}_N \circ \cdots
\circ {\cal N}_1 \circ {\cal N}_{\rm in})(\tau,x;\T)$, ${\cal N}_k(y;\T_{k}) = f(W_k y + b_k)$. Here $\T = (\T_{k})_{k=0,\cdots,N+1}$,
$\T_{k} = (W_k,b_k)$ are the neural parameters to be found, $f$ is called an activation function, which must be nonlinear, and
$N$ is the depth of the neural network.
The size of the matrices $W_k$ determines the so-called capacity: the larger it is, the more precise and time-consuming the method becomes.

Let us address important remarks about the choice of ansatz~\eqref{nnansatz}.
The boundary conditions are actually relaxed: the neural network is free to choose the best decays at $\tau \to \pm \infty$ in accordance with
the problem~\eqref{eq:BVb}.
For instance imposing a wrong decay, say $\mW \underset{-\infty}{\sim} {\rm e}^{2\tau}$ will have no effect, and the ending result will recover the correct one $\mW \underset{-\infty}{\sim} {\rm e}^{\tau}$.
The main purpose in this formulation is to avoid being attracted directly to the trivial zero solution.
That said, we observe that the precise details of the ansatz we adopt are unimportant for solving the problem, and many other formulations are possible.
Indeed, the solutions obtained do not (and must not) depend on the ansatz chosen.
This was checked many times.
As we are going to present below, the formulation we chose yields good results, but we do not claim in any way it is optimal.

In this formulation, the problem consists in finding optimal NN parameters.
Therefore, we rewrite the cost functional~\eqref{mlcost} as an optimization of $\T$ for the NN ansatz~\eqref{nnansatz} in the form
\be
\label{nncost}
{\cal C}[\T] = \int_{-T}^T \int_{x_{\rm min}}^{x_{\rm max}} 
|\dot \mW_{\rm nn} - {\cal F}[\mW_{\rm nn}]|^2(\tau,x;\T)~dx~d\tau.
\ee
We call $\T^\star = {\rm argmin}_{\T} {\cal C}[\T]$.
The NN function $W_{\rm nn}(\tau,x;\T^\star)$ represents a parametrized family of solutions from which we identify the best candidate for the eigenvalue problem by solving $\dot \mW - {\cal F}[\mW] = 0$.
This technique enables us to avoid minimizing separately many functionals for every choice of $x$.
It works well in practice provided the interval $[x_{\rm min},x_{\rm max}]$ is small enough.
The critical $x^\star$ one is looking for is therefore
\be
\label{argminx}
x^\star = {\rm argmin}_x~\log \int_{-T}^T |\dot \mW_{\rm nn}-{\cal F}[\mW_{\rm nn}]|^2(\tau,x;\T^\star)~d\tau.
\ee

Numerically, we discretize the cost functionals, and we employ the stochastic gradient descent algorithm ADAM~\cite{kingma2015adam}.
Further numerical and algorithmic details are given in the Appendix~\ref{app:NN}.

\begin{figure}
	\includegraphics[width=0.6\textwidth]{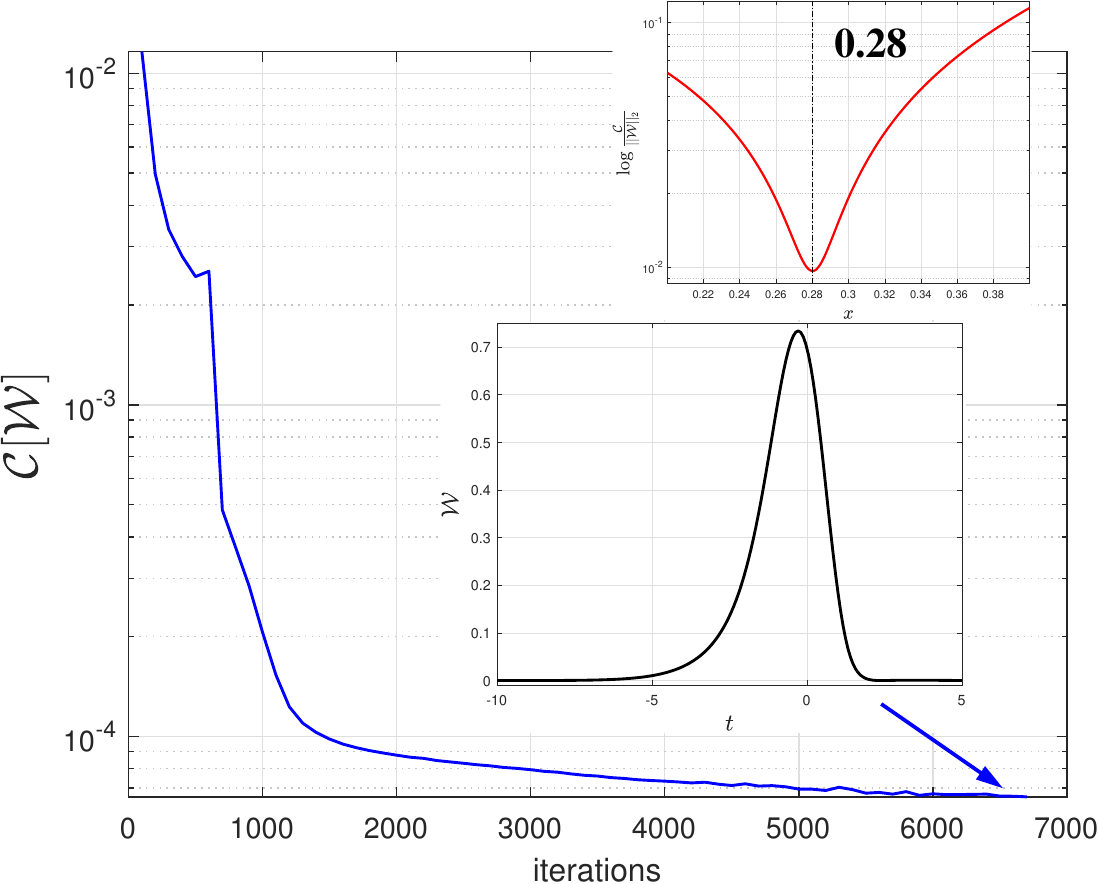}
	\caption{One-bump self-similar profile for $\lambda = 2$, $g=1$ obtained using a machine-learning approach. The background figure corresponds to the evolution of the cost functional during the stochastic gradient descent (ADAM).
		The upper panel is a measure of the relative error as a function of $x$, by plotting the expression inside the argmin of Eq.~\eqref{argminx}.
		Numerical parameters are $T=15$, $\sigma = 8$.
		The neural network has 10 hidden layers with 10 neurons/layer and swish activation functions $f$.
		Batch size is 100 $\times 50$ points for $(\tau,x)$.}
	\label{fig:8}
\end{figure}

We show in Fig.~\ref{fig:8} the results obtained after a few iterations.
Here we consider classical Sabra, \textit{i.e.} $\lambda = 2$ and $g = 1$.
The algorithmic stochastic descent is shown together with the relative error as a function of $x$ and the converged solution $\mW_{\rm nn}$.
One can observe the typical profile of the traveling wave in DG with its power-law behavior to the left and its fast decay to the right---compare it, for instance, with the profile from direct numerical integrations of Fig.~\ref{fig:2}.
Moreover, we also observe that the minimum of the cost functional is achieved in the same scaling exponent $x \approx 0.281$ obtained from the direct numerical integrations.
The expression inside the argmin in Eq.~\eqref{argminx} is shown in the insert of Fig.~\ref{fig:8}.
Those first observations attest the efficacy of the method.

\begin{figure}
	\includegraphics[width=\textwidth]{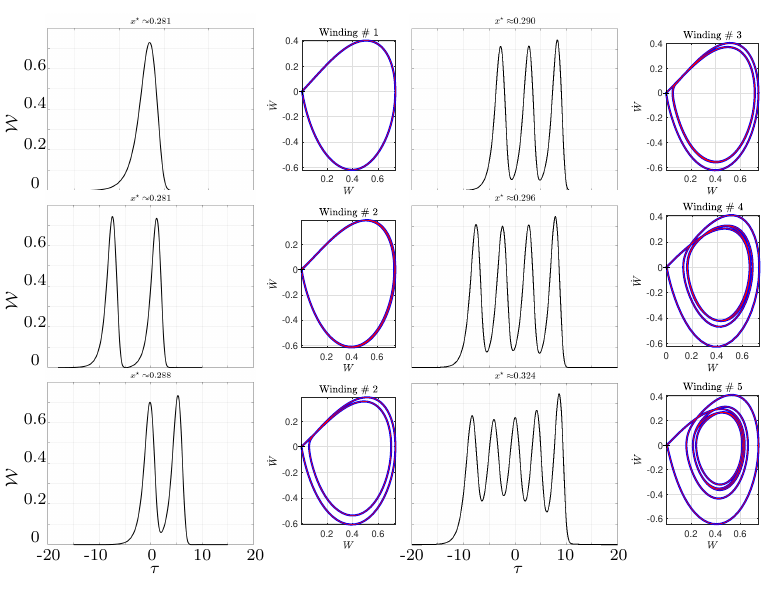}
	\caption{$N$-bumps self-similar solutions for $\lambda=2$, $g=1$, obtained using a machine-learning approach.
		The phase portraits in the plane $\mW,\dot \mW$ are displayed showing the increase of the winding number.
		As the number of bumps increases, so does the critical $x^\star$, as long as the bumps do not remain departed from each other.
	}
	\label{fig:9}
\end{figure}

\paragraph{N-bumps solutions.}
We recall that our machine-learning approach described previously relies 
on stochastic gradient descent.
Due to that, depending on the realization of the noise and/or on changing slightly the hyperparameter setup, one can obtain radically different solutions.
We show some of them in Fig.\ref{fig:9}.
In particular, one is able to obtain solutions exhibiting an arbitrary number $N$ of bumps, each one being very similar to the reference one-bump solution.
For example, one may compare the one-bump solution in Fig.\ref{fig:7} with the one- and two-bumps solutions in Fig.\ref{fig:9}.

A curious observation is the behavior of a N-bumps solution with respect to the distance between the many bumps.
When the bumps come closer to each other, the shapes and the peaks of the various bumps may differ from each other, and the global picture of the solution becomes more complex---for instance, see the five-bumps solutions in Fig.~\ref{fig:9}.

Another observation that is correlated with the distance between the bumps is the value of the scaling exponents $x_N^\star$ associated to the $N$-bumps solution.
They take values rather close to the reference one-bump exponent $x_1^\star \approx 0.281$, but more importantly they always seem to be larger than the ground (eigen)value $x^\star_{1} \leq x_N^\star$, and they approach this lower bound as the distance between the bumps increase.
We explain this observation by the following argument.
Consider the BVP~\eqref{eq:BVb} for the self-similar profile $\mW(\tau)$, and make the change of variables $\tau = \delta s$ and $\mH(s) = \delta\mW(\delta s)$.
Then, the governing equation becomes $\dot{\mH} - \delta\mH = \lambda^{-2}\mH_1\mH_2 - (1+c)\mH_{-1}\mH_{1} + c \lambda^2\mH_{-2}\mH_{-1}$, where we use the notation $\mH_j(s) = \mH(s+j)$ for the shifted functions.
In this formulation, the shifts are independent of $x$, which appears in the equation through the coefficient $\delta = (1-x) \log \lambda$.
One verifies through the energy balance that the scaling exponent $x^\star$ satisfies
\be
\label{Raylei}
1-x^\star = \frac{\mW_{\rm H}^{-1}}{\log \lambda} \frac{\displaystyle \int_{\field{R}} \mH_{-1} 
	\mH \mH_{1}~ds}{\displaystyle \int_{\field{R}} \mH^2~ds},
\ee
where $\mW_H$ is a constant given below in~\eqref{eq:constant}.
Next, we consider a $N$-bumps solution $\mH^{(N)}$ approximetely composed by the sum of $N$ one-bump solutions sufficiently departed from each other.
We write $\mH^{(N)} \approx \sum_{k=1}^N H_k$, where $H_k(s) = \mH^{(1)}(s + j_k)$ with $j_1 \ll j_2 \ll \dots \ll j_N$, such that $H_kH_l \approx 0$ if $k \neq l$.
In this case, we have $\int_\mathbb{R} \mH^{(N)}_{-1}\mH^{(N)}\mH^{(N)}_1~ds \approx N \int_\mathbb{R} \mH^{(1)}_{-1}\mH^{(1)}\mH^{(1)}_1~ds$ and $\int_\mathbb{R} \left( \mH^{(N)}\right)^2~ds \approx N \int_\mathbb{R} \left( \mH^{(1)}\right)^2~ds$.
Therefore, $x_N^\star \approx x_1^\star$.
See, for instance, the two-bumps solution with reasonably departed peaks in Fig.\ref{fig:9}, for which $x_2^\star \approx x_1^\star$ within numerical estimation.
On the other hand, if the bumps become closer to each other, the different bumps have nontrivial contributions to the nonlinearities in expression~\eqref{Raylei}, and this might lead to distinct exponents.
This is, for example, the case of the five-bumps solution displayed in Fig.\ref{fig:9}, whose exponent $x_5^\star \approx 0.324$, which is slightly bigger than $x_1^\star$.
We conjecture that the one-bump solution provides the maximal value of~\eqref{Raylei}.
We next discuss the stability of those $N$-bumps solutions.

\paragraph{Numerical simulations.}
Our machine-learning approach reveals strong nonuniqueness of self-similar blowups in Sabra.
While the classical one-bump blowup has been verified numerically to be rather stable, a natural question would be wether those multiple-bump profiles are stable or not.
We investigate this question through direct numerical integrations of DG Sabra dynamics~\eqref{eq:sabra2}.

\begin{figure}
\hspace*{-0.3cm}\includegraphics[width=\textwidth]{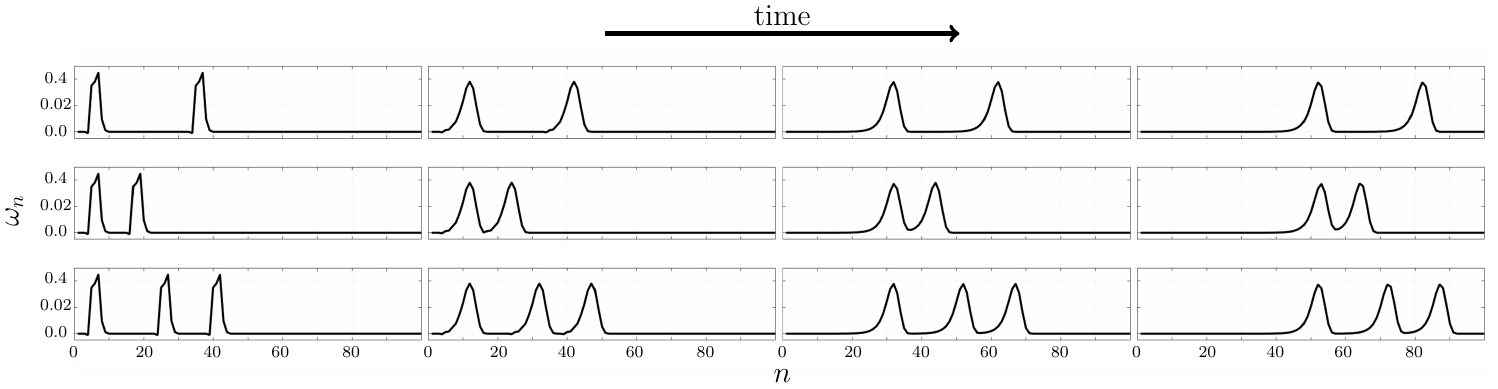}
\caption{Numerical simulations of $N$-bump self-similar solutions in the DG representation.
		Each row displays the time evolution from a different initial condition.}
\label{fig:10}
\end{figure}

We consider initial conditions exhibiting two or three bumps of the same shape, but with distinct intensities.
The initial shape of each bump is arbitrarily chosen. 
We show the time evolution of some of the solutions in Fig.~\ref{fig:10}.
We observe a relaxation from our arbitrarily chosen initial condition to the $N$-bump profiles found with machine learning---for instance, compare the numerical integrations from Fig.\ref{fig:10} to the machine learning outputs in Fig.\ref{fig:9}.
Such relaxation occurs in the first instants of evolution.
After some integration time, we observe the suppression of the $N$-bump configuration by the classical one-bump profile.
By fine tuning the intensities of the initial bumps, the profile is kept for longer times.
In principle, one could carry the $N$-bump profile for arbitrarily large times, as long as the machine precision allows the fine adjustment of the bumps initial intensities.

This whole scenario suggests that the multiple-bump solutions are saddle points of the BVP~\eqref{eq:BVb}.
The fine-tuning of parameters just guarantee closeness of the initial conditions to the stable manifold, whose contraction proportionates the observed initial relaxation.
But perturbations (even the numerical integration) may eventually disturb the system on unstable directions, and this causes the supression of the $N$-bump profile, and a subsequent convergence to the stable one-bump traveling wave.
Therefore, strictly speaking, those multiple-bump solutions do not seem to be stable in the sense of Lyapunov.
We claim, however, that more systematic studies are needed for a precise characterization of the stability of such solutions.

\section{Conclusions} 
\label{sec:4}
We investigated self-similar finite-time blowups in the inviscid Sabra shell model.
To this end, we derived the associated similarity equations and formulated them as a boundary value problem parametrized by the scaling exponent $x$.
{\color{black}Since we restricted our analysis to purely imaginary solutions, all our results also immediately hold for the GOY model.}
{\color{black}The resulting} equations involve past and future time shifts, which turns out to be difficult treating analytically.
To address this, we employed two distinct and complementary strategies.

First, we considered a perturbative approach by Taylor-expanding the shifts in the equations with respect to the bookkepping parameter $\delta = (1-x)\log \lambda$. This allowed us
to map the similarity equations to a hierarchy of dynamical systems of increasing order.
There, the blowup scenario is linked to a robust sequence of bifurcations: fixed point, Hopf bifurcation, limit cycle, and finally a homoclinic orbit, which corresponds to the desired blowup solution.
Despite the infinite-dimensional nature of the problem, we verified that the main features of the blowup are accurately described by a low-order truncation of the Sabra hierarchy.
Indeed, the $7$th order expansion already presented results indistinguishable from the theoretical predicted bifurcations and the numerically computed scaling exponents.
Similar to blowups observed in differential (Leith) models of turbulence,  blowup solutions in Sabra  emerge as codimension-1 global (homoclinic) bifurcations. Unlike Leith blowups \cite{grebenev2013self}, the high-dimensionality of the BVP revealed by the Sabra hierarchy \emph{a priori} allows for less generic bifurcations, meaning that the hypothesis of uniqueness of Sabra blowups should be studied in more depth.

Second, we employed a nonperturbative approach to directly solve the similarity equations without approximations. 
This was possible by reformulating the problem in an optimization framework, and leveraging standard machine learning techniques.
With this approach, we discovered  infinitely many distinct blowup solutions characterized by $N$-bump profiles.
While the classical one-bump profile was found to be stable, the multiple-bump profiles appeared to lack
stability, although they could still be accurately tracked using  direct numerical simulations.

We also observed that the scaling exponent of the one-bump solutions $x_\star^{(1)}$ is likely to be the minimum of all scaling exponents, \text{i.e.} $x_\star^{(N)} \geq x_\star^{(1)}$ for all $N$.
We leave the question open as to whether the multiple-bump profiles can be found  in the Sabra hierarchy.
The $N$-bump blowup solution would then be associated to multiple homoclinic orbits with winding number $N$,
but  further investigations are required to better understand the role of such structures in blow-up studies.  

Together, the two approaches reveal the nontrivial topology of self-similar blowups in the Sabra shell model.
While only one (stable) finite-time blowup is generically observed in numerical simulations, there are actually infinitely many other blowups with more complex profiles.

Their absence from prior reports might be due to their inherent instability.  The machine learning approach proves to be an excellent tool for tracking these unstable solutions and these findings open interesting perspectives for the study of more complex models.
For instance, one might wonder whether scenarios even more intricate than those described here for Sabra could arise in the 3D Euler equations or related models. It is well-known that self-similar finite-time blowups in 3D Euler cannot occur from smooth initial data with sufficiently strong vorticity decay at infinity~\cite{chae2007nonexistence}, necessitating more specific initial conditions~\cite{elgindi2021finite}. Machine learning approaches could serve as promising tools to identify not only stable blowups but also potentially unstable ones in these more complex equations.

Beyond self-similar blowups, the Dombre-Gilson scheme and its variants have revealed that cascade models could exhibit a variety of blowup scenarios~\cite{mailybaev2012renormalization}, including the possibility of a chaotic blowup~\cite{campolina2018chaotic}.
One may wishfully  ask whether combining hierarchy  and machine learning approaches could elucidate the roles of unstable similarity solutions in the emergence of those more complex  roads to blowups. We leave those interesting matters for future work.

\section*{Acknowledgements}
The authors wish to  thank  J. Bec and N. Valade for their valuable continuing discussions and debates on this topic.
C.C. was supported by the French National Research Agency (ANR project TILT ANR-20-CE30-0035),
and dedicates this work to Notre Dame de Laghet.
\appendix
\section{Local analysis}
\label{app:local}

In the following, we discuss bifurcations results for the system (\ref{eq:BVb}) written
in the dynamical system form $\dot \mW  = \mathcal F\left[ \mW \right]$. 
It is important to notice that, as such, one deals with a genuine infinite-dimensional dynamical system since the right-hand side involves delays. For convenience, we will refer
as "time" the variable $\tau$. We first briefly discuss the fixed points.

One is looking for solutions satisfying $\mathcal F\left[ \mW \right] = 0$ for all time. 
Since $\dot \mW = 0$, only the constant solutions are relevant and they 
must obey $\mW(1 + \mW(\lambda^{-2} - 1 - c + c \lambda^2)) = 0$. Since $c < 0$
(3D case) and $\lambda > 1$, it gives (\ref{eq:constant}). 

\subsubsection{Linearized dynamics and Hopf bifurcation.}\label{appH}
We consider here the linearisation of (\ref{eq:BVb}) around a fixed point
$\overline{\mW}$. The interest is that one can
obtain explicit formula. Writing $\mW =\overline{\mW} + \phi$, it is straightforward to obtain at first order in the perturbation $\phi$, the linearized dynamics
$$
\dot \phi - \phi = \overline{\mW} \left( \lambda^{-2}(\phi_{+\delta}+\phi_{+2\delta}) -(1+c) (\phi_{-\delta}+\phi_{+\delta}) + c\lambda^2
(\phi_{-2\delta} + \phi_{-\delta})
\right) ,
$$
where we use again the shorthand$\phi_{k\delta}(\tau) = \phi(\tau + (1-x) k \log \lambda)$ for the shifted functions with respect to integer $j$ multiples of $\delta = (1-x) \log \lambda$.
Applying the Fourier transform yields (with the convention $\hat \phi(\xi) = \int_{\field{R}} \phi(\tau) {\rm e}^{-i \xi \tau}~d\tau$),
\be \label{sfer}
\Gamma(\xi) \hat \phi(\xi) = 0, ~~\Gamma(\xi) = (i\xi-1) - \overline{\mW} \left( \lambda^{-2}({\rm e}^{i\delta \xi}+{\rm e}^{2i \delta\xi})
-(1+c) ({\rm e}^{-i \delta\xi}+{\rm e}^{+i \delta \xi}) + c\lambda^2 ({\rm e}^{-i\delta \xi}+{\rm e}^{-2i \delta \xi})
\right).
\ee
The case $\overline{\mW} = 0$ corresponds to $\phi(\tau) = C {\rm e}^\tau$, indicating that 0 behaves as a saddle.
Note that it is spectrally unstable since the Jacobian is the identity.
Let us turn now to the other fixed point $\overline{\mW} = 
\mW_{\rm H}$.
Its stability is dictated by the behavior of the zeros of $\Gamma$ in the complex plane. In particular, a transfer
of stability occurs when $\Im \xi = 0$ with nonzero parameter velocity (\textit{e.g.} parameter $x$).
We thus denote $\Omega = \delta \xi \in \field{R}$ and compute the real and imaginary parts of $\Gamma$, namely
\be
\Re \Gamma = -1 -\overline{\mW}  \left((\lambda^{-2} -2(1+c) + c\lambda^2) \cos  \Omega + (\lambda^{-2}+c \lambda^2)
\cos 2 \Omega \right),
\ee
and
\be
\Im \Gamma = \delta^{-1} \Omega  -\overline{\mW} (\lambda^{-2} -c \lambda^2) (\sin \Omega + \sin 2 \Omega).
\ee
The real part is quadratic in term of $\cos \Omega$. We now take $\overline{\mW} = \mW_{\rm H}$ (see (\ref{eq:constant}))
and solve $\Re \Gamma = 0$. Let $\rho_{1,2}$ be the roots of
\be
\rho^2 + \left(\frac12 -R\right) \rho + \frac{R}{2}-1 = 0,~~R := \frac{c+1}{c \lambda^2 + \lambda^{-2}}.
\ee
We are looking for a root such that $|\rho|\leq 1$. The two roots are $2 \rho= R-\frac12 \pm \sqrt{ R^2 -3 R + \frac{17}{4} }$,
where the expression inside the squareroot is always positive and larger than 2. The condition for having $|\rho| \leq 1$ 
imposes either $R \leq 1$ or $R \geq \frac13$. As a consequence, there is always a root such that $|\rho | \leq 1$.
In fact the two roots satisfy this property provided $R \in [\frac13,1]$. It turns out that this scenario {\it never occurs when
$c < 0$}. Otherwise, it would implies not one, but at least two bifurcations, together with possible codim-2 bifurcations.

One can then proceed with $\Omega = \arccos \rho$, namely $\xi = \gamma^{-1} \arccos \rho$
and the imaginary part $\Im \Gamma$ gives $\xi =\mW_{\rm H} (\lambda^{-2} -c \lambda^2) (\sin \Omega + \sin 2 \Omega)= \mW_{\rm H} (\lambda^{-2} -c \lambda^2) \sqrt{1-\rho^2} (1+2\rho)$. 
One is interested in the dependence on the power-law exponent $x$. It appears in the term $\delta$ only in the real part  of
$\Gamma$ (since
$\mW_{\rm H}$ does not depend on $x$). In other words, the necessary condition for a bifurcation to occur
is
\be
1-x_{\rm Hopf}= \frac{(\lambda^2-1) (1-c \lambda^2)}{(1-c \lambda^4)\log \lambda}
\frac{\arccos \rho}{(1+2\rho) \sqrt{1-\rho^2}},
\ee
with frequency
\be
\xi_{\rm Hopf} = \frac{1-c \lambda^4}{(\lambda^2-1)(1-c \lambda^2)} (1+2\rho) \sqrt{1-\rho^2}.
\ee

\begin{figure}
\includegraphics[width=\textwidth]{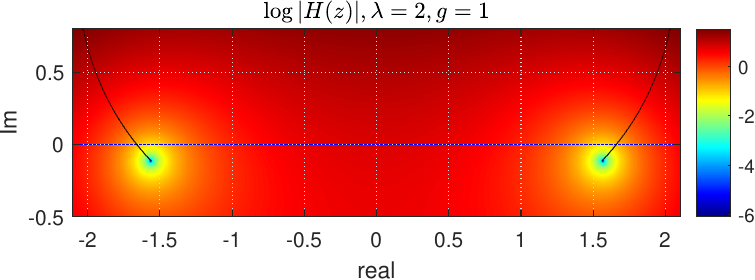}
\caption{Poles (black curves) of the function $\xi \to \Gamma(\xi)$, $\xi \in \field{C}$ when $x$ varies. The
norm $\log |\Gamma|$ in the background is computed at the ending value of $x$. The Hopf bifurcation corresponds
to the crossing of the black curves with the real axis.}
\end{figure}

\subsubsection{The limit $\lambda \to 1^+$.}
 Let us write $\lambda = 1+\epsilon$ and assume that
$c = -\lambda^{-g}$. One easily obtains $1-x \sim \frac{2 \arccos \rho}{(1+2 \rho) \sqrt{1-\rho^2}}$ and
$R \sim \frac{g}{4-g}$ so that $\rho$ is well defined except maybe for $g=4$. In fact when $g \to 4$, the roots
are $\{-\infty,\frac12 \}$. We obtain in this particular case $1-x_{\rm Hopf} \to 1 - \frac{2 \pi}{3\sqrt{3}}$.
The frequency term behaves as $\xi \sim \frac{1}{2\epsilon} (1+2\rho) \sqrt{1-\rho^2}$. It is quite remarkable that
although the fixed point $\mW_{\rm Hopf} \sim \frac{1}{4 \epsilon}$ and the frequency are pushed to infinity when $\epsilon \to 0$, a bifurcation still occurs at a well defined value of $x$ which depends on $g$ using the same formula than before but with $R =
\frac{g}{4-g}$ .

\section{The Sabra hierarchy}\label{app:hierarchy}
We here provide a technical derivation of the Sabra hierarchy, and comment on the  choice of  $\delta = (1-x)\log\lambda$ as a natural bookkeeping parameter. The hierarchy proposed here is justified only in the case $|\delta|<1$ and
converges to the continuous problem modulo an ad-hoc rescaling.

\subsubsection{Technical derivation.}
We derive the  Sabra hierarchy \eqref{eq:BVPm} by Taylor  expanding the shifted functions as 
\be
	\mW_k(\tau) = \sum_{i=0}^M  \dfrac{k^i\delta^i}{i!}\mW^{(i)}(\tau) + O(\delta^{M+1}),
\ee
which in turn yields for the $\mF$ -functional
\be
	\label{eq:Fn}
	 \mF[\mW]  =  \mW^{(0)}(\tau) + \dfrac{1}{2}\sum_{k=0}^M\delta^k \sum_{i+j=k} \sigma_{ij} \mW^{(i)}(\tau)\mW^{(j)}(\tau)+O(\delta^{M+1})
\ee
in terms of the symmetric coefficients $\sigma_{ij}$ defined by Eq.~\eqref{eq:coeffs}.
Neglecting the terms $O(\delta^{M+1})$ in Eq.~\eqref{eq:Fn}, the Sabra BVP \eqref{eq:BVb} truncates into the $M^{th}$ order system
\be
	\label{eq:SabraFn}
	 \mW^{(1)}(\tau)= \mF_M[\mW] ,\quad \mF_M[\mW] =  \mW^{(0)}(\tau) + \dfrac{1}{2}\sum_{k=0}^M\delta^k \sum_{i+j=k} \sigma_{ij} \mW^{(i)}(\tau)\mW^{(j)}(\tau).
\ee
Besides, the boundary conditions \eqref{eq:BVc} become
\be 
\forall k,\beta \ge 0\quad  \; \mW^{(k)}(\tau) \underset{-\infty}{\propto} e^\tau \to 0, \quad \mW^{(k)}(\tau) \underset{\infty}{=} o\left(e^{-\beta\tau}\right) \to 0.
\ee
Writing $X_k = \delta^k \mW^{(k)}$ and $\bX(\tau) = (X_0,\cdots,X_{M-1})$,  Eq.~\eqref{eq:SabraFn} recasts into the first order system
 \be
\label{eq:Xtau}
\left\lbrace
\begin{split}
 \delta\; \dot X_k(\tau) & = X_{k+1},\quad  0\le k \le M-2,\quad 
	 \sigma_{0M} \delta X_0 \dot X_{M-1}(\tau)  = \mG_M\left[\bX\right],\\
& \quad \mG_M\left[\bX\right]= - X_0 + \dfrac{X_1}{\delta}  -\dfrac{1}{2} \sum_{k=0}^{M-1}\sum_{\substack{i+j=k\\ 0 \le i,j \le k}}  \sigma_{ij} X_i X_{j}- -\dfrac{1}{2} \sum_{\substack{i+j= M\\ 1 \le i,j \le M-1}}  \sigma_{ij} X_i X_{j},
\end{split}
\right.
\ee
with boundary conditions $\bX(\tau) \underset{\pm \infty}\to 0$.
System.~\eqref{eq:Xtau} features a term $X_0$ on its left-hand-side (lhs)  and is therefore singular. To desingularize it, we introduce the new time variable $\theta=\int_0^\tau \frac{d\tau'}{\delta X_0(\tau')} $ as prescribed by
Eq.~\eqref{eq:theta} --  in short   $d\theta=(\delta X_0)^{-1} \; d\tau$. This yields the Sabra hierarchy \eqref{eq:BVPm}.
For the boundaries, the change of variable $\tau \to \theta$ obviously prescribes
$\theta(\tau) \to_{-\infty} -\infty$.
 Does it obviously prescribe $\theta(\tau) \to \infty$ as $\tau\to \infty$?
One must assume that $\mW > 0$ ($X_0 > 0$) for all $\tau$, if not the change of variable $\theta$ is not well-defined.
The boundary condition at $+\infty$ is $X_0 = o(e^{-\beta \tau})$. It is equivalent to write
$ X_0 e^{\beta \tau} = g(\tau) \to 0$ at $+\infty$ with $g > 0$. Then, there exist $\tilde \tau$ and a constant $c$
 such that $g(\tau) < 1$ for all $\tau \geq \tilde \tau$ and $g(\tau) > c > 0$ for all $\tau \leq \tilde \tau$ so that $\theta = \int_0^{\tilde \tau} \frac{e^{\beta s}}{g(s)}~ds + \int_{\tilde \tau}^\tau  
\frac{e^{\beta s}}{g(s)}~ds \geq cst + \int_{\tilde \tau}^\tau e^{\beta s}$, therefore $\theta \to \infty$ when $\tau \to \infty$.
\color{black}

\subsubsection{The asymptotics $\delta\to 0 $. }
The asymptotics $\delta \to 0$ describes the case    $x \to 1^-$ or the formal limit $\lambda \to 1$.  In this asymptotics, the Sabra BVP  \eqref{eq:BVb} becomes local and the hierarchy \eqref{eq:Xtau} reduces to the case  $ M= 2$.  In this situation, the Sabra BVP and the hierarchy reduce to
\be
\dot \mW=\mW +\dfrac{1}{2}\sigma_{00}\mW^2,\quad \sigma_{00} = -2\mW_{\rm H}^{-1}, \quad \text{(+ BC)}
\ee
yielding the explicit (logistic) solution
$W(\tau) = \dfrac{1}{\mW_{\rm H}^{-1} +C^{-1}e^{-\tau}}$. The logistic solution satisfies the vanishing boundary condition only for $\tau \to -\infty$, and hence is not a solution to the Sabra BVP. It converges to the fixed point $\mW_{\rm H}>0$ for $ \tau \to \infty$. Heuristically, we however expect the true solution to the BVP to be close to the logistic solution, a feature indeed observed in the non-perturbative optimization approach. This motivates to consider variations in the small parameter $\delta = (1-x) \log \lambda$ to match the ultraviolet boundary condition, and provides a physical rational for the  Taylor expansion in $\delta$.

\section{Machine-learning approach}\label{app:NN}
We provide here more details on the numerical simulations of~\S\ref{sec:3}, which uses a neural network parametrisation.
First of all, the cost functional~\eqref{nncost} must be discretized.
We consider a set of $N_\tau \times N_x$ independent and identically distributed (i.i.d.) random points $(\tau,x)$ in the functional domain ${\cal D}_{\tau,x} = [-T,T] \times [x_{\rm min},x_{\rm max}]$ with uniform distribution ${\cal U}([-T,T]) \times {\cal U}(x_{\rm min},x_{\rm max})$.
Then, we use a Monte-Carlo integral approximation for the cost functional
\be\label{discost}
{\cal C}[\T] \approx \frac{1}{N_\tau} \frac{1}{N_x} \sum_{k=1}^{N_x} \sum_{j=1}^{N_\tau} 
\left| \dot {\cal W}_{\rm nn}(\tau_{j,k},x_{j,k};\T) - {\cal F}[{\cal W}_{\rm nn}(\tau_{j,k},x_{j,k};\T)] \right|^2,
\ee
where ${\cal W}_{\rm nn} = {\cal W}(\tau,x;\T)$ establishes a neural network with parameters $\T$ to be found---recall~\eqref{nnansatz}.
The gradients of~\eqref{discost} are found by automatic differentiation (backpropagation).
The algorithm follows the simple rules:
\begin{itemize}
\item[$\bullet$] {\bf Hyperparameter setup}:
choose $T,\sigma$ and build some NNs with input dimension equals to 2 (for $\tau$) and output dimension equals to $1$.
Choose $N_\tau,N_x$ and a learning rate $\eta$.
\item[$\bullet$] {\bf Main loop}: REPEAT until satisfactory convergence (${\cal C} \ll 1$)
\begin{itemize}
\item[1)] Draw $N_\tau \times N_x$ i.i.d. points in ${\cal D}_{\tau,x}$, denoted by ${\bf m} = (\tau_{ij},x_{ij})_{i,j}$;
\item[2)] Compute $\nabla_{\T} {\cal C}[\T]({\bf m})$ from (\ref{discost}) by automatic differentiation;
\item[3)] Update $\T$:  $\T \mapsto \T - \eta \nabla_{\T} {\cal C}$. We employ the stochastic descent algorithm ADAM~\cite{kingma2015adam};
\item[4)] Monitor every few steps ${\cal W}$ and ${\cal C}$. 
\end{itemize}
\end{itemize}
In practice, we use a feedforward NN with 10 hidden layers, with 10 neurons/layer (the weights matrices $W$ are of size $10 \times 10$), and we use a swish activation function $f$~\cite{ramachandran2017searching}.
The training rate is $\eta = 10^{-3}$ (ADAM) decreased to $\eta = 10^{-4}$ at the end of the descent. The typical parameters
in the simulations are $15 \leq T \leq 20$, $\sigma = 8$, $N_\tau = 100$, $N_x = 50$.
These parameters are not fine-tuned and can be modified.
More powerful setups are possible, but they are not discussed here.

\bibliography{biblio}
\end{document}